\documentstyle[epsfig]{elsart}
\begin{document}

\newcommand{\re}{\mathop{\mathrm{Re}}}
\newcommand{\im}{\mathop{\mathrm{Im}}}
\newcommand{\I}{\mathop{\mathrm{i}}}
\newcommand{\D}{\mathop{\mathrm{d}}}
\newcommand{\E}{\mathop{\mathrm{e}}}

\def\lambar{\lambda \hspace*{-5pt}{\rule [5pt]{4pt}{0.3pt}} \hspace*{1pt}}

{\Large  DESY 14-025}

{\Large  March 2014}

\bigskip

\bigskip

\bigskip

\begin{frontmatter}

\journal{Nucl. Instrum. and Methods A}

\date{}

\title{
Prospects for CW and LP operation of the European XFEL in hard
X-ray regime}

\author{R.~Brinkmann,}
\author{E.A.~Schneidmiller,}
\author{J.~Sekutowicz,}
\author{M.V.~Yurkov}

\address{Deutsches Elektronen-Synchrotron (DESY),
Notkestrasse 85, D-22607 Hamburg, Germany}

\begin{abstract}
The European XFEL will operate nominally at 17.5 GeV in SP (short pulse) mode with 0.65 ms long bunch train and
10 Hz repetition rate.
A possible upgrade of the linac to CW (continuous wave) or LP (long pulse) modes with a corresponding
reduction of electron beam energy
is under discussion since many years. Recent successes in the dedicated R\&D program allow to forecast a technical
feasibility of such an
upgrade in the foreseeable future. One of the challenges is to provide sub-\AA ngstrom FEL operation in CW and LP modes.
In this paper we perform a preliminary analysis
of a possible operation of the European XFEL in the hard X-ray regime in CW and LP modes with the energies of 7 GeV and 10 GeV,
respectively. We consider lasing in the baseline XFEL undulator as well as in a
new undulator with a reduced period.
We show that, with reasonable requirements on electron beam quality,
lasing on the fundamental will be possible in sub-\AA ngstrom regime.
As an option for generation of brilliant photon beams at short wavelengths we also
consider harmonic lasing that has recently attracted a significant attention.
\end{abstract}

\end{frontmatter}

\baselineskip 20pt

\clearpage

\section{Introduction}

Successful operation of X-ray free electron lasers (FELs) \cite{flash,lcls,sacla}, based on
self-amplified spontaneous emission (SASE) principle \cite{ks-sase},
opens up new horizons for photon science.
The European XFEL \cite{euro-xfel-tdr} will be the first hard X-ray FEL user facility based on superconducting
accelerator technology, and will
provide unprecedented average brilliance of photon beams. The XFEL linac will operate nominally at
17.5 GeV in a burst mode with up to 2700 bunches within a 0.65 ms long bunch train and 10 Hz repetition rate.
Even though the RF pulses are much longer than those available at X-ray FEL facilities,
based on normal conducting accelerators, in the context of this paper we will call this SP (short pulse)
mode of operation.

In order to cope with high repetition rate within a pulse train,  special efforts are being made to develop fast X-ray
instrumentation \cite{detector}. Still, many user experiments would strongly profit from increasing distance between X-ray pulses while
lengthening pulse trains and keeping total number of X-ray pulses unchanged (or increased). Such a regime would require an
operation of the accelerator with much longer RF pulses (LP, or long pulse mode), or even in CW (continuous wave) mode as a limit.
A possible upgrade of the XFEL linac to CW or LP modes with a corresponding
reduction of electron beam energy is under discussion since many years \cite{brinkmann}.
Recent successes in the dedicated R\&D program \cite{jacek} allow to forecast a technical
feasibility of such an upgrade in the foreseeable future.

One of the main challenges of CW upgrade is to provide sub-\AA ngstrom FEL operation which is, obviously,
more difficult with lower electron energies. One can consider improving the electron beam quality as well as
reducing the undulator period as possible measures. An additional possibility is a
harmonic lasing \cite{murphy,hg-2,kim-1,mcneil,parisi,sy-harm}
that has recently attracted a significant attention \cite{sy-harm,sy-harm-flash}.
Harmonic lasing can extend operating range of an X-ray FEL facility and provide brilliant photon beams of high
energies for user experiments. In this paper we briefly discuss all these possibilities.
Note that the goal of this paper is not to define precisely the parameter range for operation of the
European XFEL after CW upgrade, but to stimulate further intensive studies aiming at definition of optimal regimes.

\section{CW upgrade of the linac}

A possible upgrade of the XFEL linac to CW or LP modes holds a great potential for a further improvement of X-ray FEL user operation,
including a more comfortable (for experiments) time structure, higher average brilliance, improved stability etc. The drawbacks are
a somewhat smaller peak brilliance and a reduced photon energy range, both due to a lower electron beam energy. Both disadvantages
can, however, be minimized by an improvement of the electron beam quality and application of advanced FEL techniques.
Moreover, one can
keep a possibility to relatively quickly switch between SP and CW modes thus greatly improving the flexibility
of the user facility.

For a CW upgrade of the linac, the following main measures will be needed \cite{jacek,jacek-fel14}:

i) Upgrade of the cryogenic plant with the aim to approximately double its capacity;

ii) Installation of new RF power sources: compact Inductive Output Tubes (IOTs);

iii) Exchange of the first 17 accelerator modules by the new ones
(including a larger diameter 2-phase helium tube, new HOM couplers etc.)
designed for operation in CW mode with a relatively high gradient
(up to 16 MV/m). This ensures that the beam formation system (up to the last
bunch compressor) operates with a similar energy profile as it does in SP mode.
Then 12 old accelerating modules are relocated to the end of the linac;

iiii) Installation of a new injector generating a high-brightness electron beam in CW mode.

\begin{table}[b]
\caption{Main assumptions for CW/LP upgrade of  the XFEL facility.}
\bigskip

\footnotesize

\begin{tabular}{ l l l}
\hline \\
Parameter &   Unit   & Value \\
\hline \\
{\bf Beam} & & \\
Maximum charge per bunch   &	[nC]  &	0.5  \\
Time between subsequent bunches &	[µs]  & 	4  \\
Average current  &	[mA]  & 	0.125    \\
Maximum beam energy in CW mode  &	[GeV]  &	7.8  \\
{\bf Main Linac} & & \\
Number of 8-cavity cryomodules  &	-   &	96  \\
1.8K cryogenic dynamic load per cryomodule   &	[W]  &	16  \\
1.8K cryogenic static load per cryomodule    &	[W]  &	4  \\
$Q_0$ of cavities 	&  -   &	$2.8 \times 10^{10}$  \\
Maximum $E_{\mathrm{acc}}$ for CW mode   &	[MV/m]  &	7.3   \\
$Q_{\mathrm{load}}$ of input coupler  &  -	&  $2 \times 10^7$ \\
Mean RF-power per cavity  &	[kW]  & 	$< 2.5$  \\
Assumed microphonics: (peak-peak)/2  &	[Hz]  & 	32  \\
{\bf Injector Section up to 2 GeV}  & & \\
Number of 8-cavity cryomodules in linacs L0 / L1 / L2  &	-   &	1/ 4/ 12  \\
Maximum $E_{\mathrm{acc}}$ for CW mode in linacs L0 / L1 / L2	 &  [MV/m]   &	16 /11/ 15    \\
1.8K total cryogenic load per cryomodule in linacs L0 / L1 / L2	 &  [W]  &	91 / 45 / 80  \\
$Q_0$ of cavities &	 -  &	$2.5 \times 10^{10}$ \\
\hline \\
\end{tabular}

\label{tab:param}
\end{table}

The first two items can be realized in a straightforward way; the third one is based on the steady progress of the
TESLA technology \cite{tesla} and is not particularly challenging. Until recently the main uncertainty
was connected with the absence of CW injectors providing a sufficient
quality of electron beams. However, last year there was an experimental demonstration of small emittances
(for charges below 100 pC) at a CW photoinjector using a DC gun followed
immediately by acceleration with superconducting cavities \cite{cornell}. The measured parameters are already sufficient
for considering this kind of injector as a candidate for CW upgrade of the XFEL linac (although the operation would be
limited to low charge scenarios). As an alternative one can consider a superconducting RF gun that can potentially produce also
larger charge bunches with low emittances (the progress reports can be found in \cite{srfgun-fel14,srfgun-ipac14})
or even a normal conducting RF gun \cite{berkley-gun}.
In the latter case a special regime can, in principle, be organized when a continuous sequence of short RF pulses
is used \cite{vogel} instead of powering the gun in true CW mode.

In this paper we do not present a comprehensive technical description of the CW upgrade, it will
be published elswhere \cite{jacek-det}. Here we only summarize some technical details in Table~\ref{tab:param}.

To predict a possible electron energy range in CW and LP modes, and to test the XFEL cryomodules in these regimes, a series of
measurements is being performed at DESY \cite{jacek,jacek-fel14}.
The measurements demonstrate stable behavior of the modules in these regimes,
and allow to conservatively predict that the energy can reach 7 GeV in CW mode, and 10 GeV in LP mode
with 35\% duty factor \cite{jacek,jacek-fel14}.
Recent measurements \cite{jacek-privat} of a cryomodule equipped with large grain Nb cavities and improved HOM couplers
demonstrated even better performance, and allow for more optimistic forecasts (as it is reflected in Table 1).
Moreover, all these measurements have been done with
pre-series XFEL cryomodules which have not yet reached an ultimate performance. In other words, one can hope for
higher electron energies after CW upgrade. Nevertheless, in this paper we conservatively consider electron energy
range between 7 GeV and 10 GeV.

\section{Harmonic lasing}

Apart from the standard regime of the FEL operation, namely lasing at the fundamental wavelength to saturation,
in this paper we will also consider harmonic lasing as an option for reaching short wavelengths.
Harmonic lasing in single-pass high-gain FELs \cite{murphy,hg-2,kim-1,mcneil,parisi,sy-harm}
is the radiative instability at an odd harmonic of the
planar undulator developing independently from lasing at the fundamental. Contrary to nonlinear harmonic
generation (which is driven by the fundamental in the vicinity of saturation),
harmonic lasing can provide much more intense, stable, and narrow-band FEL beam
if the fundamental is suppressed. The most attractive feature of saturated harmonic lasing
is that the brilliance of a harmonic is comparable to that of the fundamental. Indeed, a good estimate for the
saturation efficiency is $\lambda_{\mathrm{w}}/(h L_{\mathrm{sat}})$, where $\lambda_{\mathrm{w}}$ is the undulator period,
$L_{\mathrm{sat}}$ is the saturation length, and $h$ is harmonic number. At the same time, the relative rms bandwidth
has the same scaling. If we consider lasing at the same wavelength on the fundamental and on a harmonic (with the retuned undulator
parameter $K$), transverse coherence properties are about the same since they are mainly defined by emittance-to-wavelength ratio.
Thus, also the brilliance is about the same in both cases. In many cases, however, the saturation length for harmonics can be
shorter than that of the fundamental at the same wavelength. As a consequence, for a given undulator length one can reach
saturation on harmonics at a shorter wavelength.
It was shown in a recent study \cite{sy-harm} that the 3rd and even the 5th harmonic lasing in X-ray FELs is much more robust
than usually thought, and can be widely used at the present level of accelerator and FEL technology.

For a successful harmonic lasing the fundamental mode must be suppressed.
A possible method to disrupt the fundamental without affecting the
third harmonic lasing was suggested in \cite{mcneil}: one can
use $2\pi/3$ (or $4\pi/3$) phase shifters between undulator modules. It was found out, however, that this method is
inefficient in the case of a SASE FEL (see \cite{sy-harm,sy-harm-flash} for more details). One can consider an alternation of
$2\pi/3$ and $4\pi/3$ phase shifters \cite{parisi} but this variation of the method works also not sufficiently well in many practical
cases. In \cite{sy-harm} another modification of the phase shifters method was proposed which
suggests piecewise distribution of phase shifters $2\pi/3$ and $4\pi/3$, and works better in practical situations \cite{sy-harm-flash}.
One can also consider a random distribution of phase shifters $2\pi/3$ and $4\pi/3$ \cite{huang} although one needs
a very larger number of phase shifters in this case.
The suppression of the fundamental by using a spectral filter in a chicane installed
between two parts of the undulator was also proposed in \cite{sy-harm}. The conclusion in \cite{sy-harm} was that the best way
of suppression is
the combination of phase shifters and a filter.

In the context of CW operation the use of the filter may have limitations due to power losses in it (even though the filtering is done
well below FEL saturation). Thus, the bunch repetition rate might be limited depending on charge and
wavelength. In this case one can use a new method for suppression \cite{suppression}:
switching between the 5th and the 3rd harmonics. As we will see in the
next Section, their performance is comparable in the considered parameter range. Thus, the following trick is possible.
Imagine, we aim at lasing at 1 \AA. We tune the first part of the undulator to the resonance with 5 \AA, so that we are interested
in the 5th harmonic lasing. The fundamental and the third harmonic are suppressed by the piecewise combination of (some of the)
phase shifters $2\pi/5$,
$4\pi/5$, $6\pi/5$, and $8\pi/5$ such that they stay well below saturation in the first part of the undulator. Then, in the
second part we reduce parameter $K$ such that the resonance at 3 \AA \ is achieved. Now the fifth harmonic from the first part
continues to get amplified as the third harmonic (while the first and the third ones are off resonance). The fundamental in the
second part is suppressed with the help of piecewise distribution of phase shifters $2\pi/3$ and $4\pi/3$. If necessary, one can
later switch back to a resonance with 5 \AA, and so on.
Also, a use of many pieces with 5th and 3rd harmonic lasing without phase shifters might be possible.
Moreover, the scheme can be generalized to the case of even higher harmonics.
More detailed description of the method as well as numerical simulations
will be published elsewhere \cite{suppression}. Note also that the FEL bandwidth in this scheme is defined by the 5th harmonic
lasing, while the saturation power - by the third harmonic lasing, i.e. the spectral power and brilliance will be higher than
in the case of lasing only on one harmonic or only on the fundamental. This is similar to the concepts of the
harmonic lasing self-seeded FEL \cite{sy-harm,hlss} and pSASE \cite{psase,pizdase}.

\section{Lasing in the baseline undulator}

In the following we will consider the range of beam energies from 7 to 10 GeV, assuming that the former can be achieved in CW
mode, and the latter - in LP mode with about 35\% duty factor \cite{jacek}.
The hard X-ray undulators SASE1 and SASE2 of the European XFEL have 4 cm period, and the largest K-value of 3.9 is achieved
at the gap of 10 mm. The net magnetic length of the undulator is 175 m. Our task is to define the range of achievable
photon energies depending on electron beam quality. We consider lasing on the fundamental as well as on the 3rd and the 5th
harmonics. The formulas from \cite{sy-harm} are used to calculate the saturation length.
We assume that the peak current is 5 kA in all
considered cases (different compression scenarios we leave for future studies).
We optimize beta-function in the undulator for the
shortest gain length. However, when the optimum beta is smaller than 15 m (which we assume as a technical limit), we
set it to 15 m.

\begin{figure}[tp]

\epsfig{file=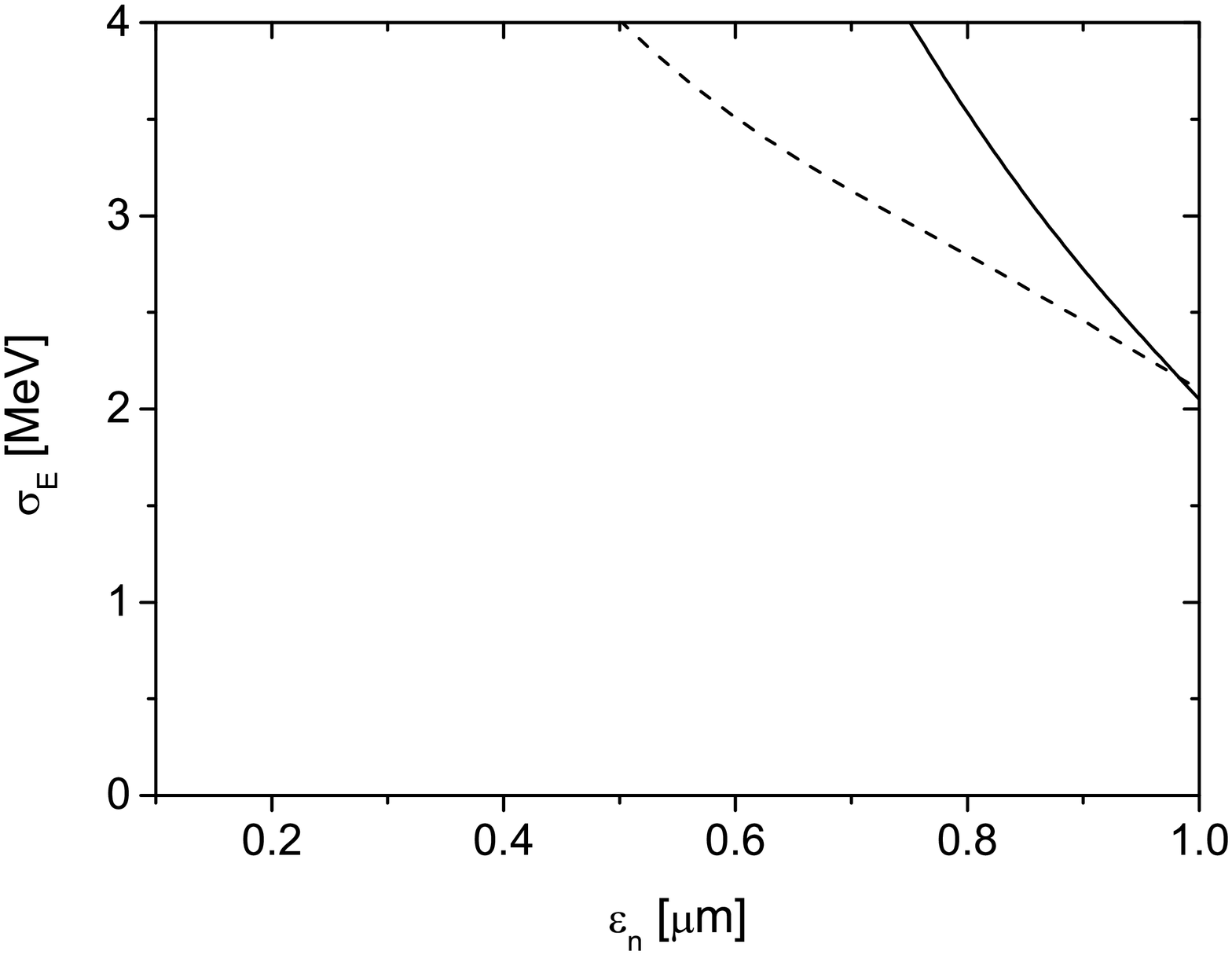,width=0.5\textwidth}

\vspace*{-56mm}

\hspace*{0.5\textwidth}
\epsfig{file=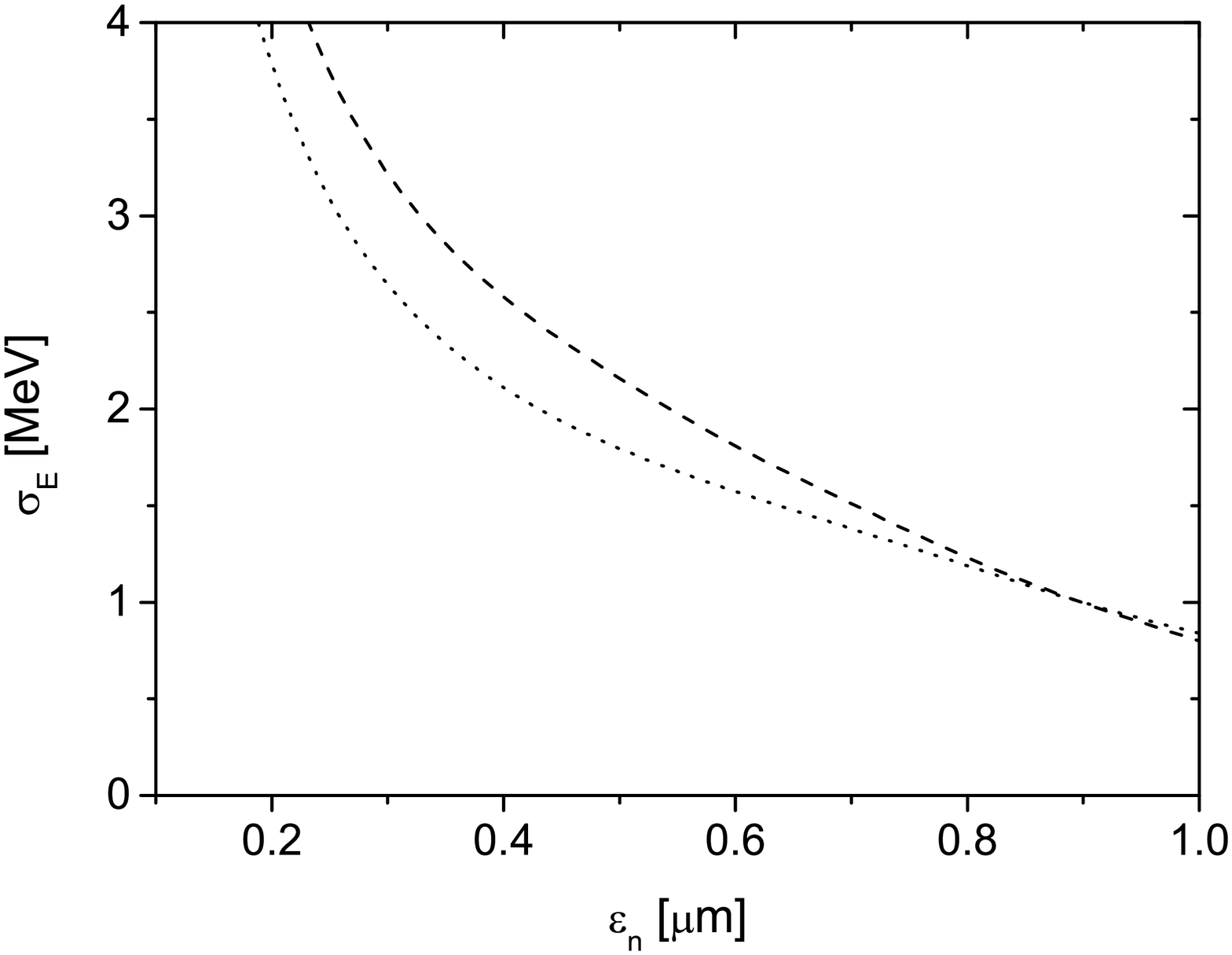,width=0.5\textwidth}

\epsfig{file=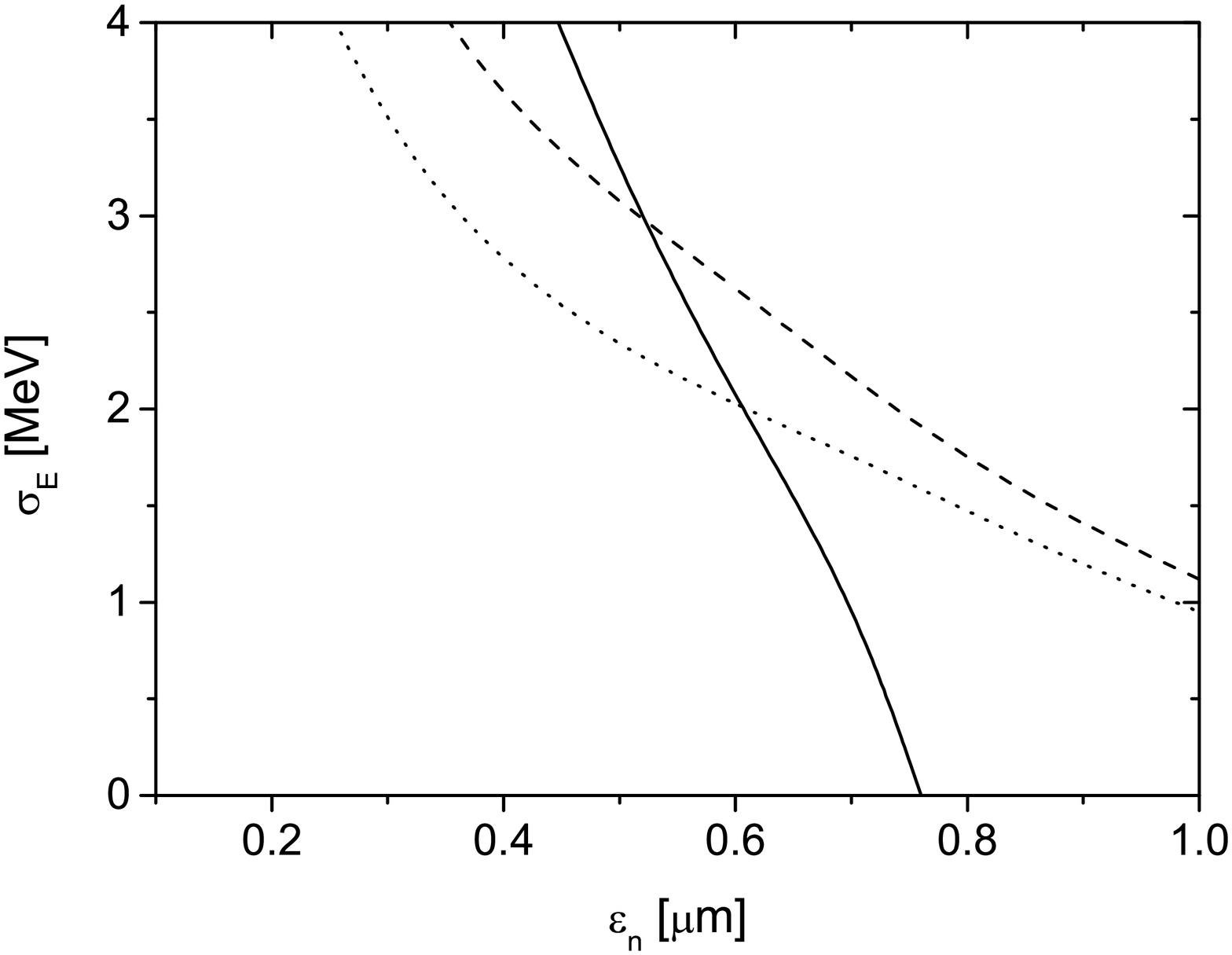,width=0.5\textwidth}

\vspace*{-56mm}

\hspace*{0.5\textwidth}
\epsfig{file=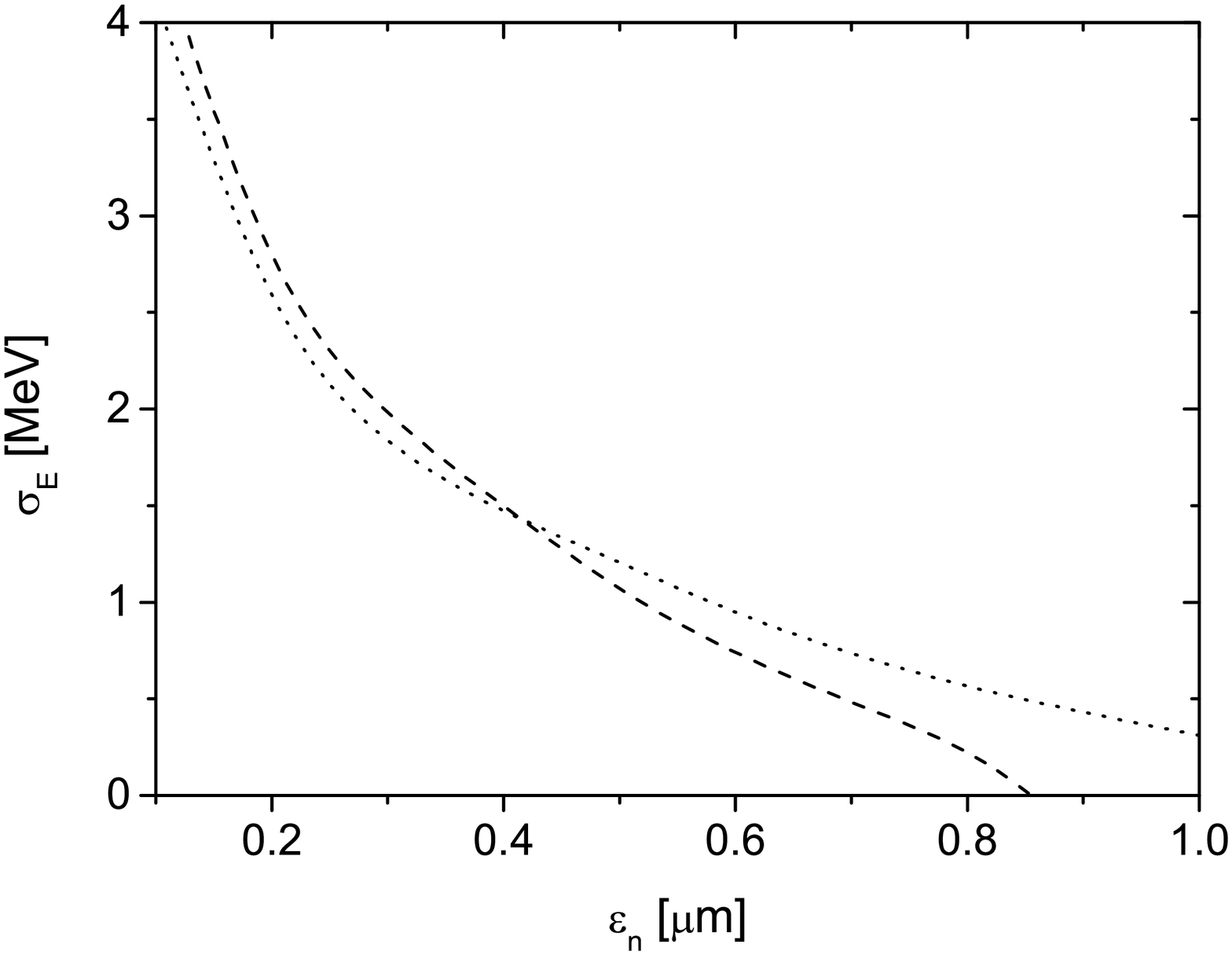,width=0.5\textwidth}

\epsfig{file=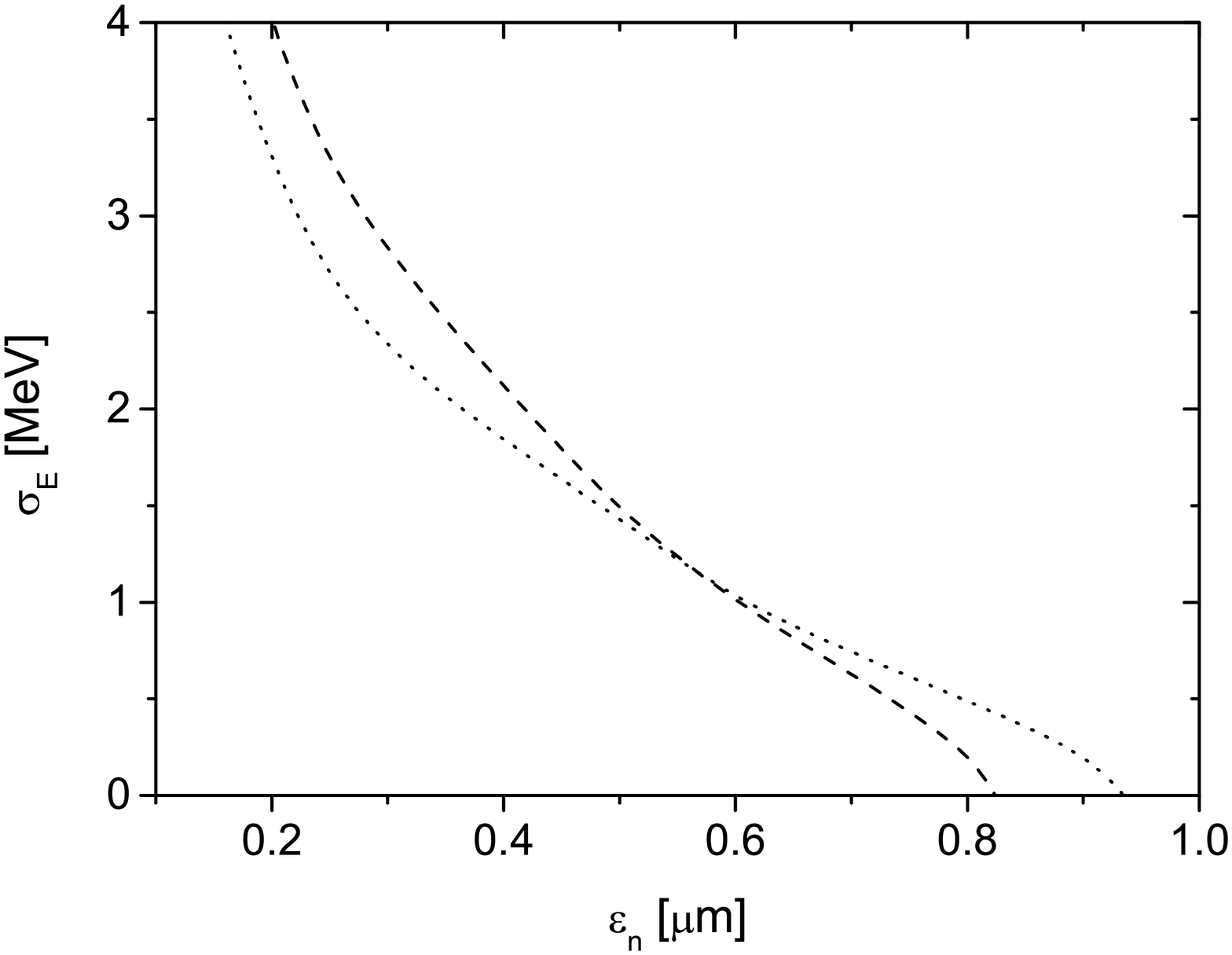,width=0.5\textwidth}

\vspace*{-56mm}

\hspace*{0.5\textwidth}
\epsfig{file=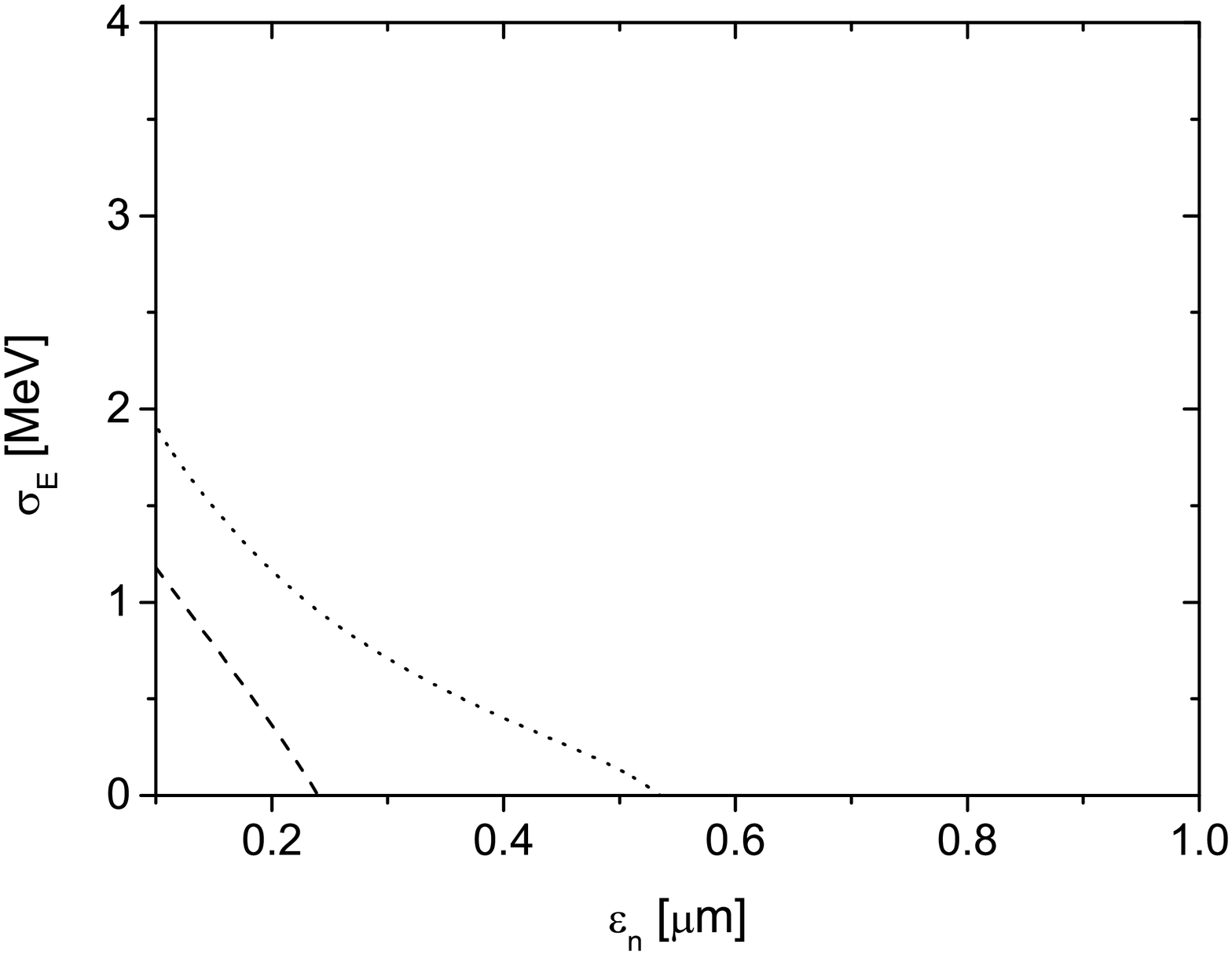,width=0.5\textwidth}

\vspace*{2cm}

\caption{\footnotesize Energy spread versus normalized emittance for which the saturation in SASE1 undulator is possible at 1 \AA
(upper graphs), 0.75 \AA (middle graphs), and 0.5 \AA (lower graphs). The
beam energy is 10 GeV (left column) and 7 GeV (right column), and the peak current is 5 kA. Beta-function is optimized for the
highest gain when the optimum is larger than 15 m, otherwise beta-function is 15 m.
Solid, dash, and dot curves correspond to the 1st, the 3rd, and the 5th harmonic lasing, correspondingly.
}

\label{fig:em-es}

\end{figure}

In Fig.~\ref{fig:em-es} we present energy spread versus normalized emittance for which the saturation is
possible at 1 \AA, 0.75 \AA, and 0.5 \AA \ for the electron energies of 7 and 10 GeV.
The corresponding photon energy range is 12.4--24.8 keV. In the case of 10 GeV the lasing at 1 \AA \ is not possible on the fifth
harmonic because K is not sufficiently large. However, lasing to saturation on the fundamental and on the 3rd harmonic is
possible practically for any reasonable beam quality. Resonance at 0.5 \AA \ cannot be achieved on the fundamental, but lasing
to saturation on the 3rd and on the 5th harmonics is possible for a sufficiently bright electron beam. In the case of 7 GeV the
resonance at 1 \AA \ and at shorter wavelengths is not possible on the fundamental. However, lasing to saturation on the 3rd and
the 5th harmonics in sub-\AA ngstrom regime is possible
(although at 0.5 \AA \ it would require extremely bright electron beams).

\begin{figure}[tp]

\epsfig{file=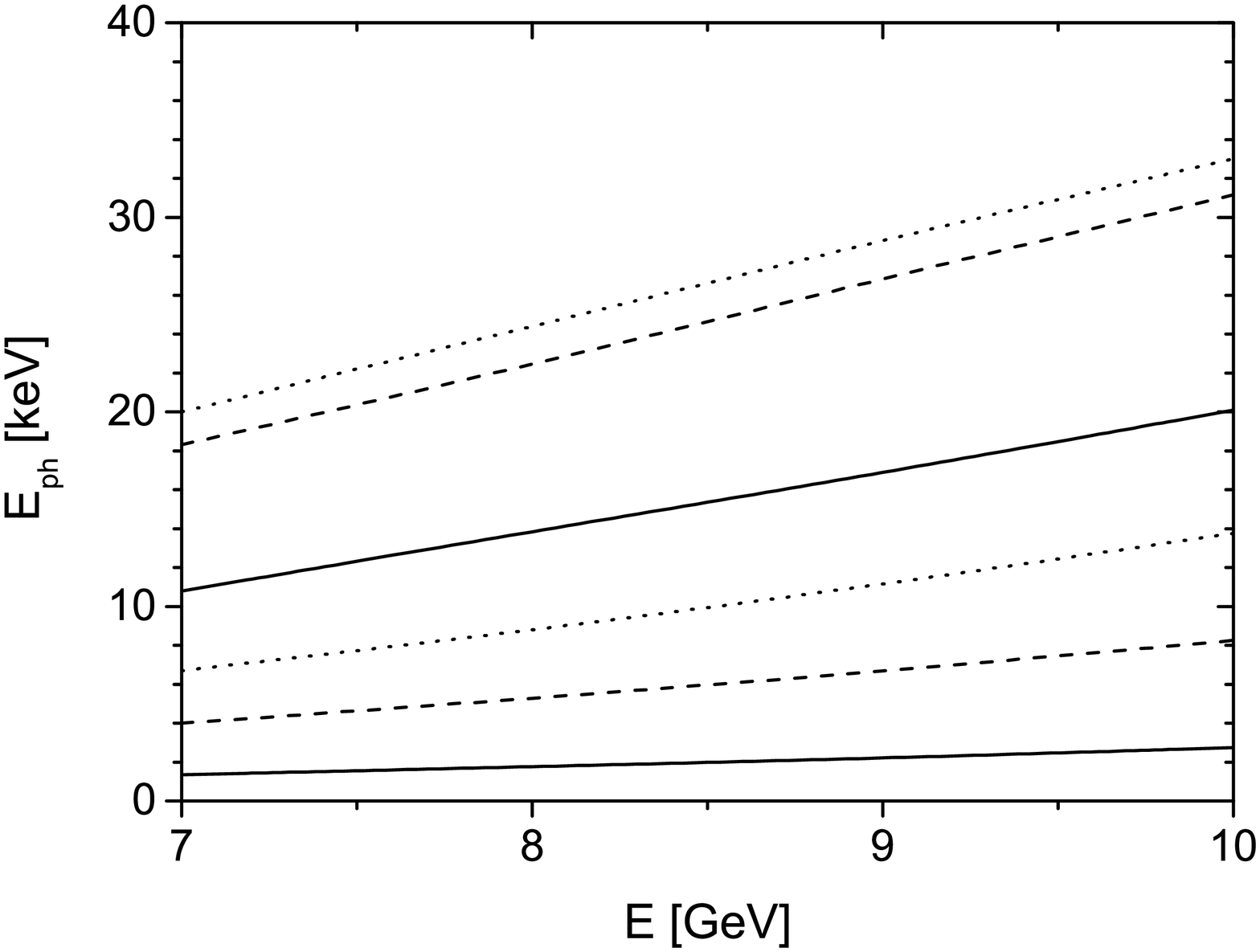,width=0.5\textwidth}

\vspace*{-56mm}

\hspace*{0.5\textwidth}
\epsfig{file=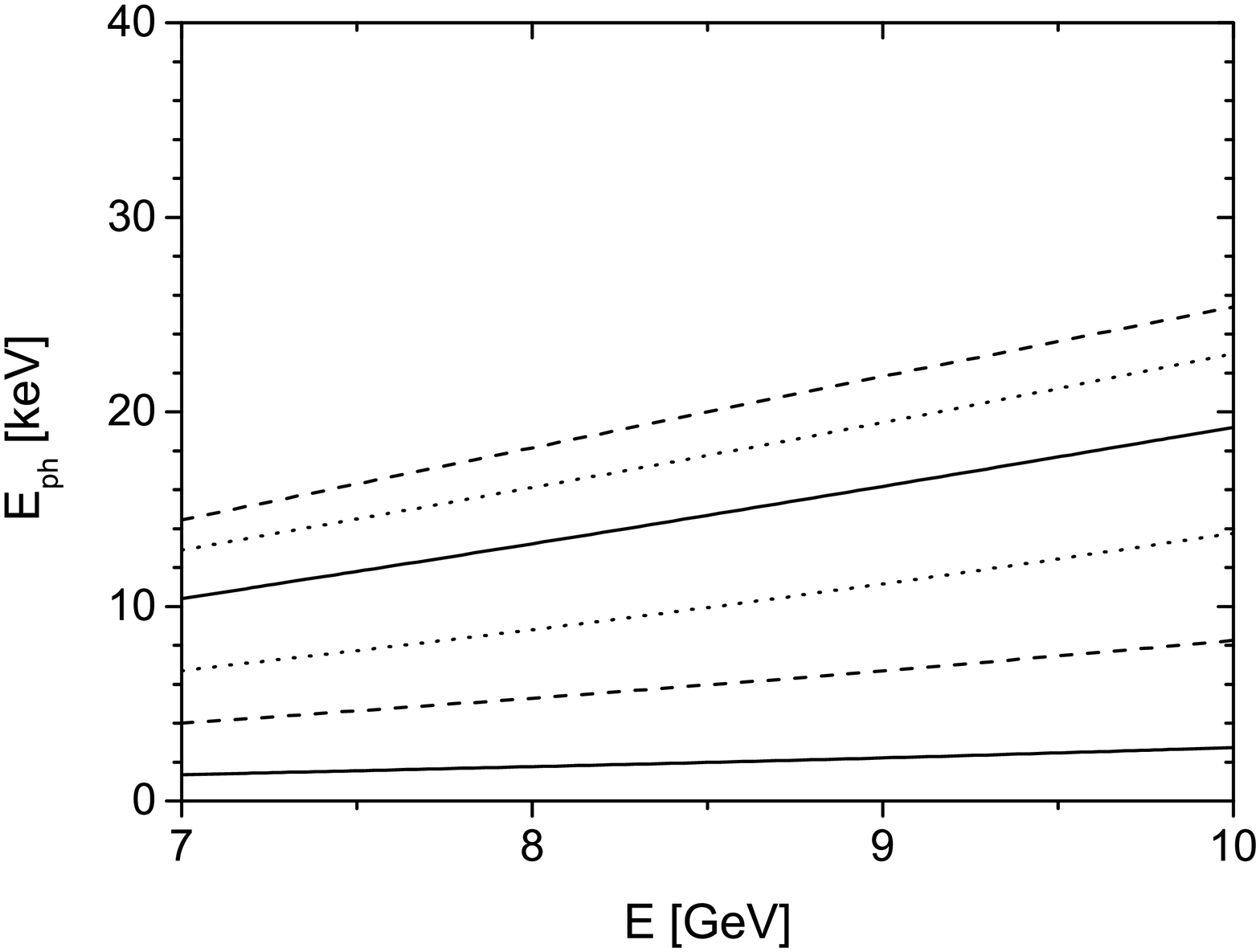,width=0.5\textwidth}

\epsfig{file=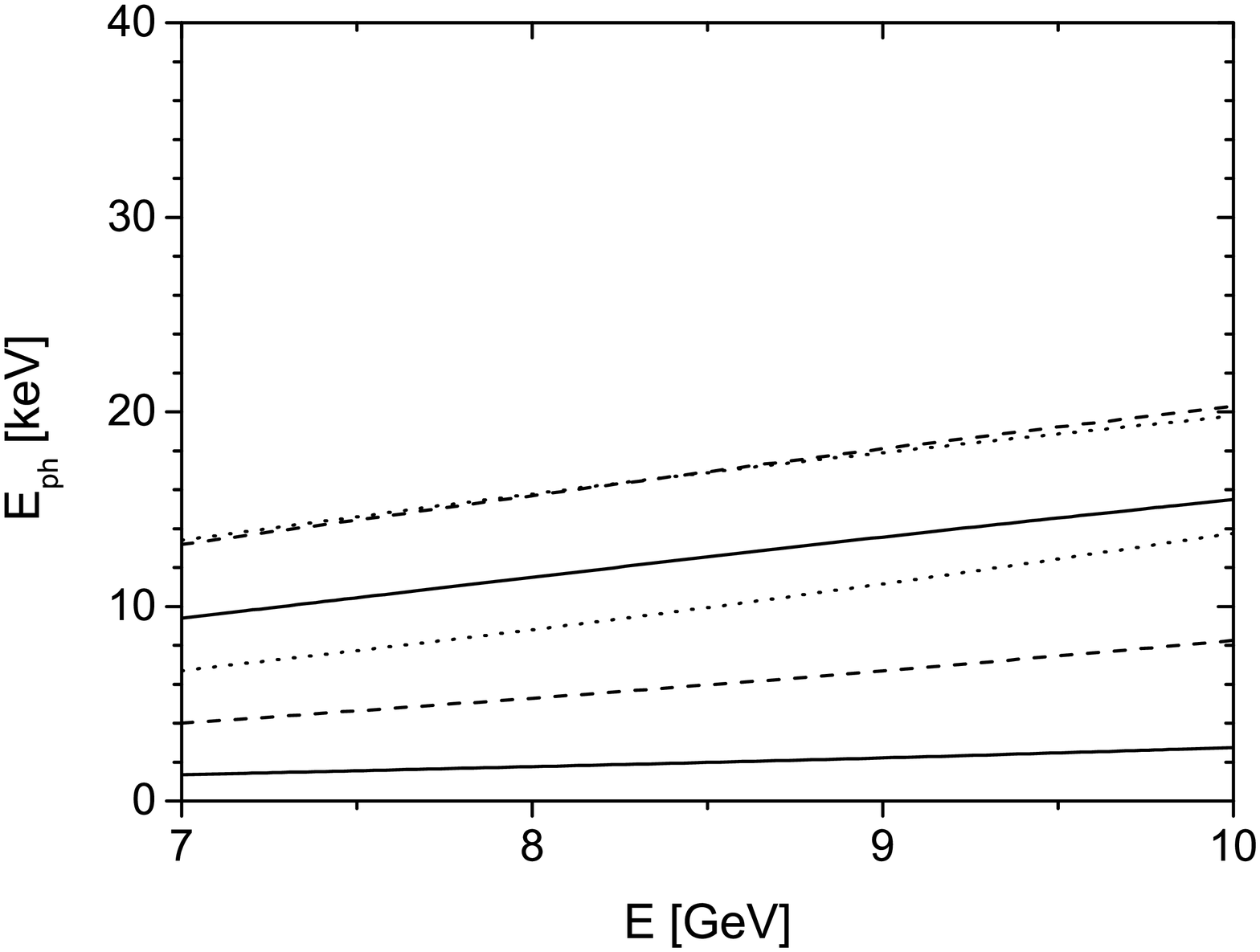,width=0.5\textwidth}

\vspace*{-56mm}

\hspace*{0.5\textwidth}
\epsfig{file=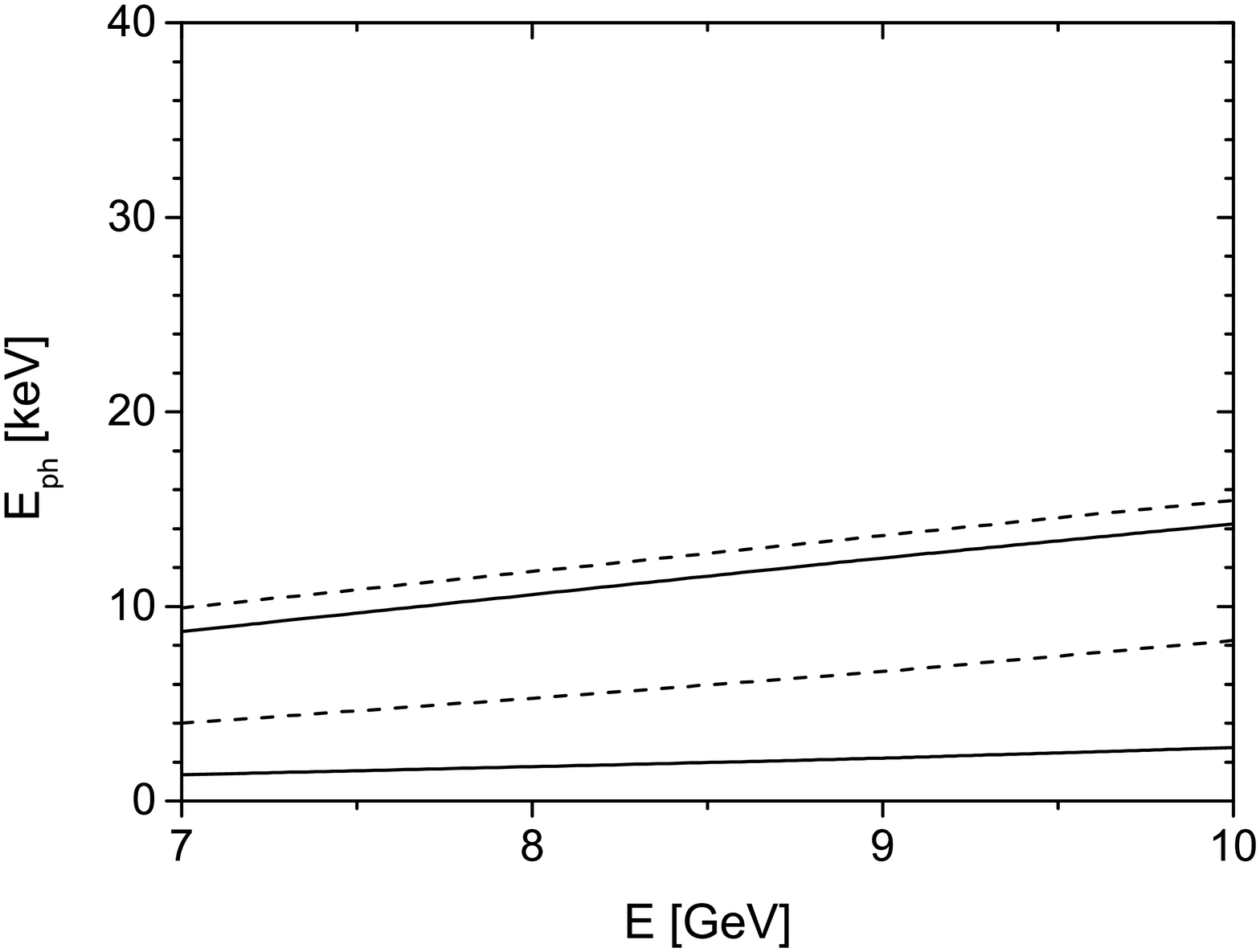,width=0.5\textwidth}

\vspace*{2cm}

\caption{\footnotesize
Range of photon energies accessible with the fundamental (between the solid lines), the third harmonic (between the dash lines),
and the fifth harmonic (between the dot lines). Upper left: $\epsilon_{\mathrm{n}}=0.4 \ \mu$m, $\sigma_{\cal{E}} = 1$ MeV,
upper right: $\epsilon_{\mathrm{n}}=0.4 \ \mu$m, $\sigma_{\cal{E}} = 2$ MeV,
lower left: $\epsilon_{\mathrm{n}}=0.8 \ \mu$m, $\sigma_{\cal{E}} = 1$ MeV,
lower right: $\epsilon_{\mathrm{n}}=0.8 \ \mu$m, $\sigma_{\cal{E}} = 2$ MeV.
The undulator period is 4 cm, maximum $K$ is 3.9, the peak current is 5 kA. Beta-function is optimized for the
highest gain when the optimum is larger than 15 m, otherwise beta-function is 15 m.
}

\label{fig:ph-el-4cm}

\end{figure}

We can also calculate photon energy range that can be achieved in the considered electron energy range depending on electron beam
quality. In Fig.~\ref{fig:ph-el-4cm} we present the results for four different combinations of slice emittance and energy
spread ranging from 0.4 $\mu$m and 1 MeV (upper left plot) to 0.8 $\mu$m and 2 MeV (lower right plot). One can see that harmonics
have a significant advantage over the fundamental only if the electron beam is bright enough. One can notice that, for example,
lasing on the 5th harmonic is not possible for the most pessimistic parameter set.

Finally, let us note that the baseline undulator with a relatively large period and large K value has an advantage of a big
tunability range, as one can see from Fig.~\ref{fig:ph-el-4cm}. Moreover,
one can keep the usual operation range after switching back to the SP mode.
Another advantage is that keeping a relatively large gap (10 mm) is favorable
in the context of CW operation with a high average power of the electron beam.

\section{Undulator with a shorter period}

When electron energy is decreased, a natural step towards reaching short wavelengths is a choice of an undulator with a
shorter period. An obvious disadvantage of such a solution is that the tunability range is reduced. In order to keep it large,
one should reduce an undulator gap thus increasing maximum K value. There are, however, limitations due to stronger wakefields and
a necessity to keep a large enough aperture for transportation of a high average power beam. In this Section we consider an
undulator with the period of 3 cm and a slightly reduced gap of 8 mm. The maximum K value, that is achieved at a closed gap,
equals 3.

\begin{figure}[tp]

\epsfig{file=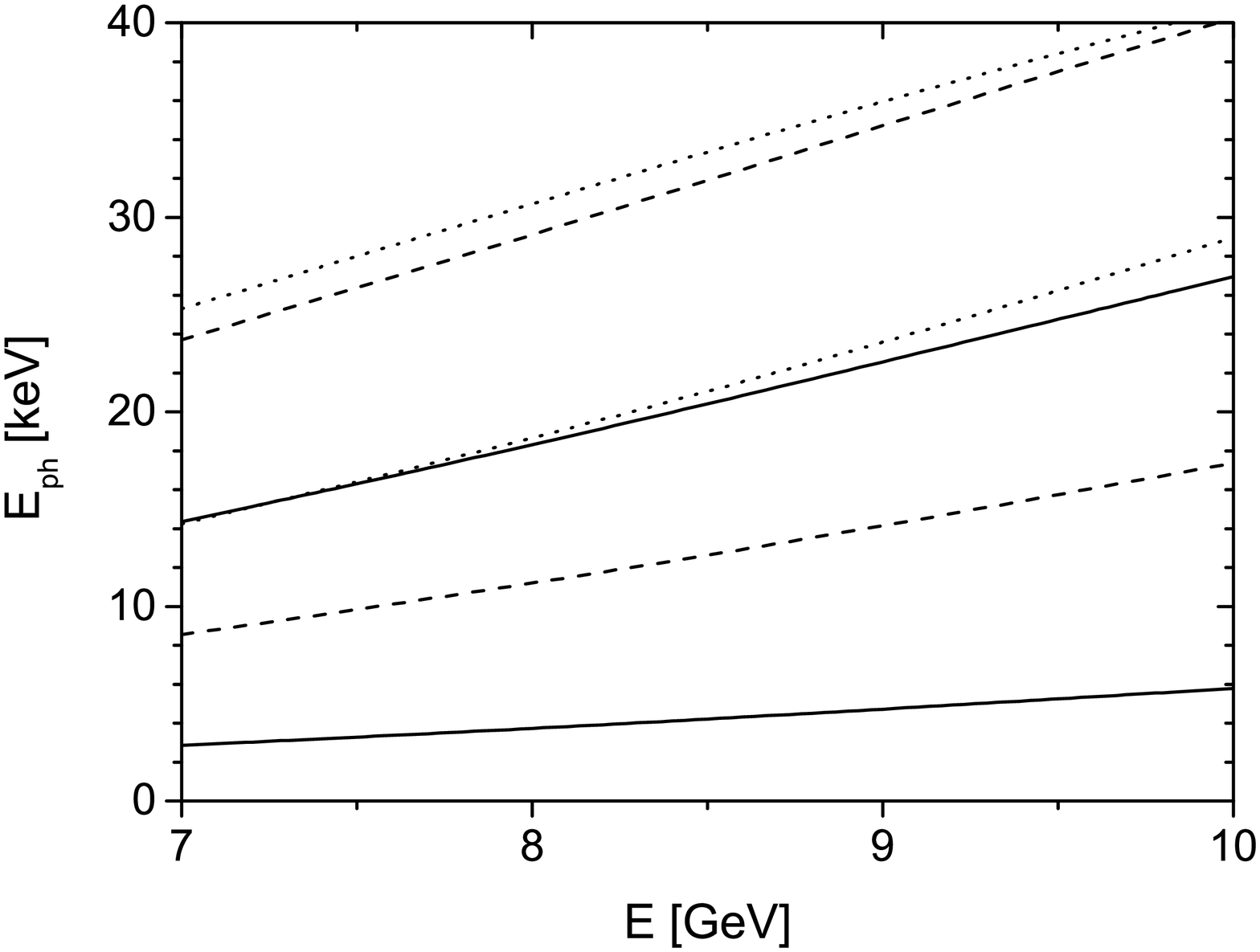,width=0.5\textwidth}

\vspace*{-56mm}

\hspace*{0.5\textwidth}
\epsfig{file=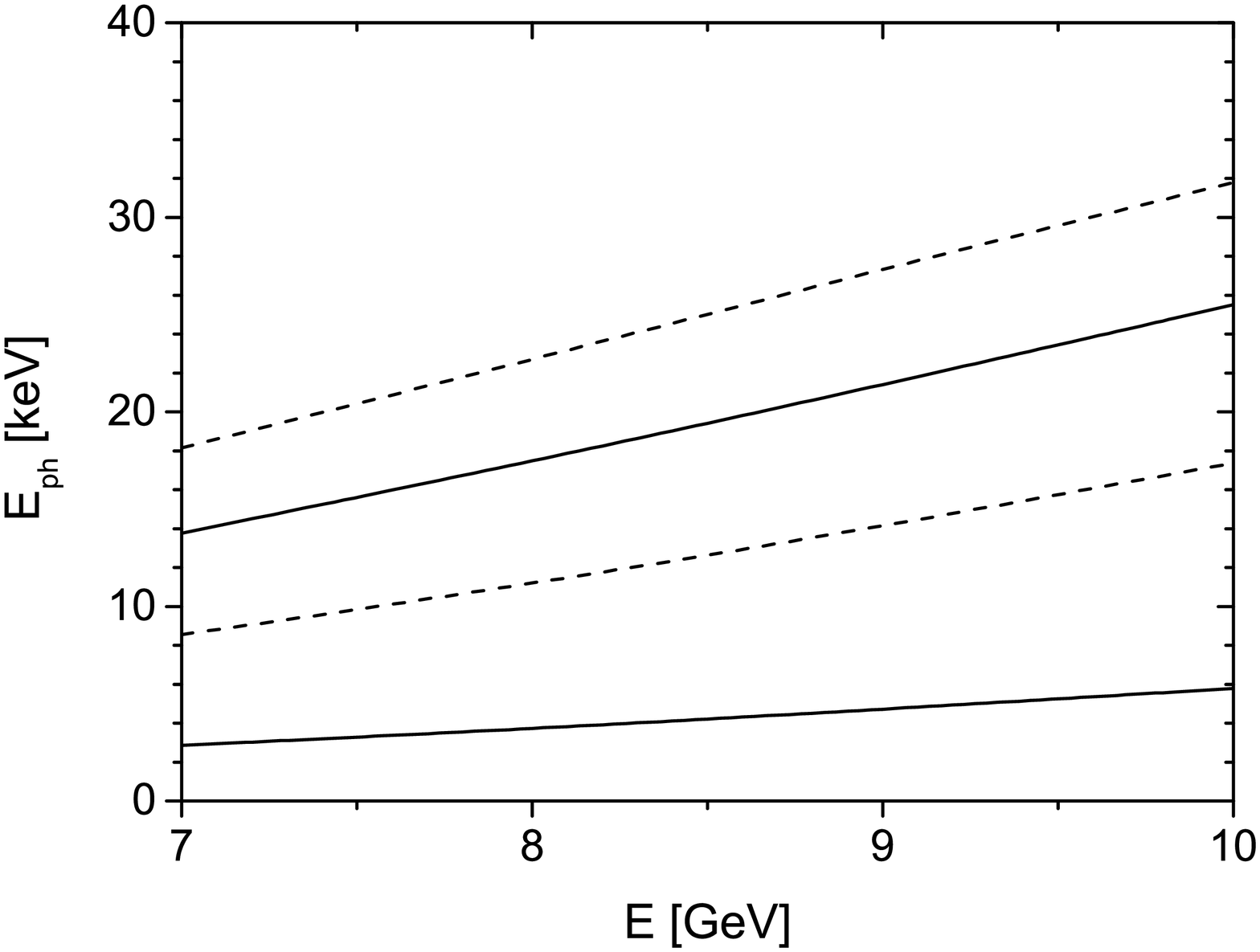,width=0.5\textwidth}

\epsfig{file=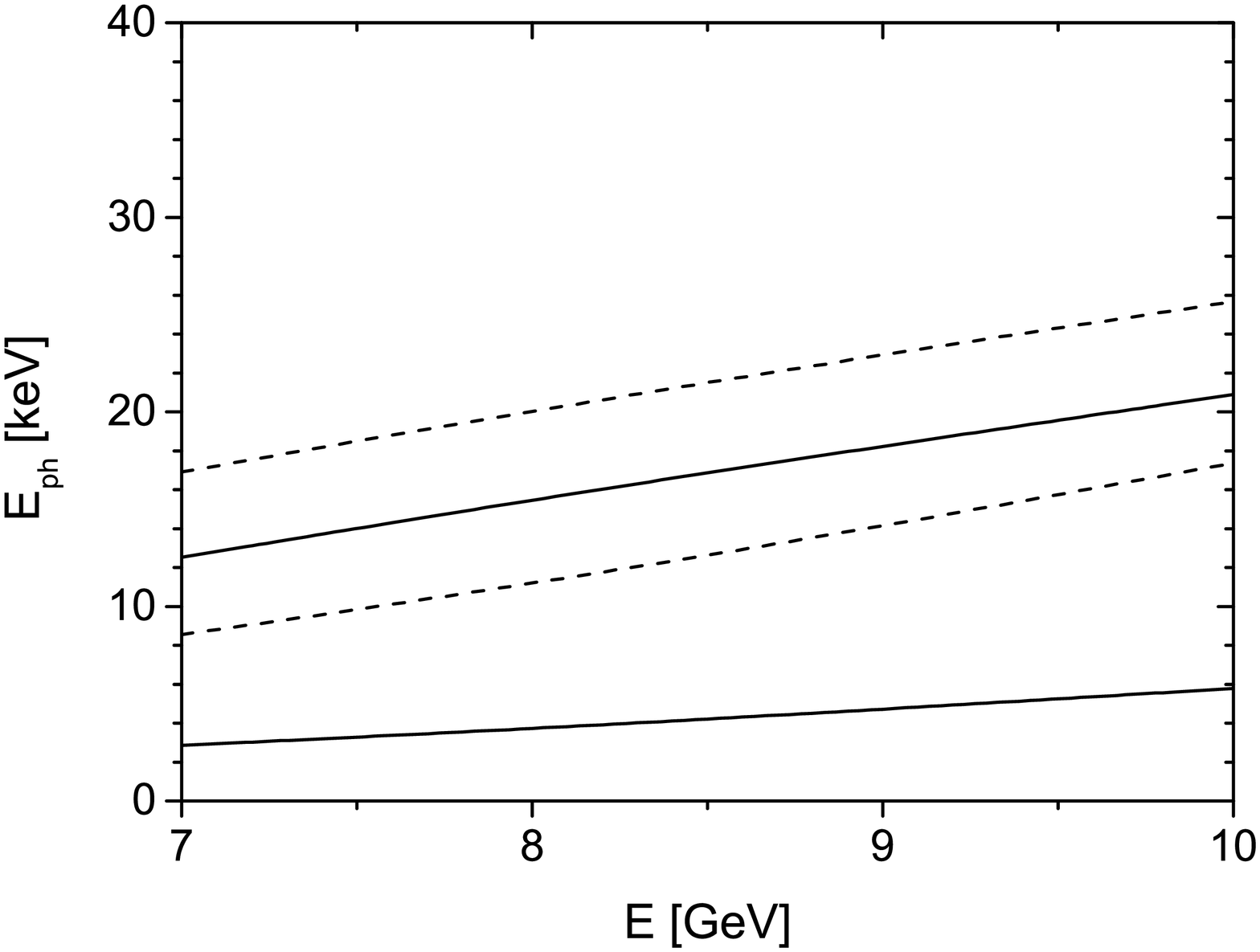,width=0.5\textwidth}

\vspace*{-56mm}

\hspace*{0.5\textwidth}
\epsfig{file=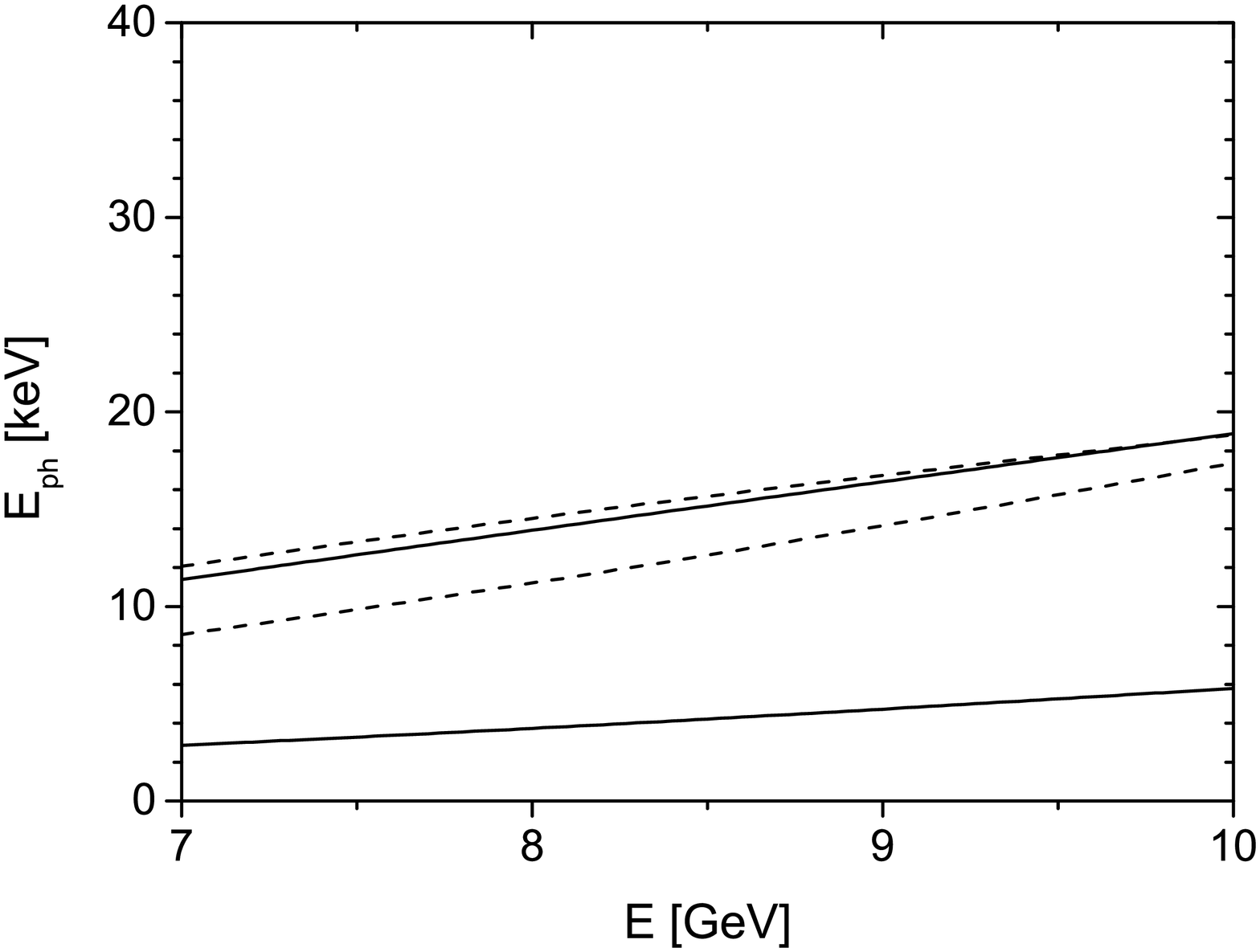,width=0.5\textwidth}

\vspace*{2cm}

\caption{\footnotesize
Range of photon energies accessible with the fundamental (between the solid lines), the third harmonic (between the dash lines),
and the fifth harmonic (between the dot lines). Upper left: $\epsilon_{\mathrm{n}}=0.4 \ \mu$m, $\sigma_{\cal{E}} = 1$ MeV,
upper right: $\epsilon_{\mathrm{n}}=0.4 \ \mu$m, $\sigma_{\cal{E}} = 2$ MeV,
lower left: $\epsilon_{\mathrm{n}}=0.8 \ \mu$m, $\sigma_{\cal{E}} = 1$ MeV,
lower right: $\epsilon_{\mathrm{n}}=0.8 \ \mu$m, $\sigma_{\cal{E}} = 2$ MeV.
The undulator period is 3 cm, maximum $K$ is 3, the peak current is 5 kA. Beta-function is optimized for the
highest gain when the optimum is larger than 15 m, otherwise beta-function is 15 m.
}

\label{fig:ph-el-3cm}

\end{figure}

In Fig.~\ref{fig:ph-el-3cm} we show the photon energy range for different sets of emittance and energy spread,
ranging from 0.4 $\mu$m and 1 MeV (upper left plot) to 0.8 $\mu$m and 2 MeV (lower right plot). Comparing Fig.~\ref{fig:ph-el-3cm}
with Fig.~\ref{fig:ph-el-4cm}, we can observe that there is a shift towards higher photon energies and a reduction of the range
(in other words, an increase of the highest achievable photon energies is smaller than an increase of the lowest ones).
One can also notice that the fifth harmonic lasing is only possible for the most optimistic parameter set.
A possible advantage of using a new undulator is that one can foresee more phase shifters already at the design stage. In particular,
one can simultaneously introduce phase shifters between undulator modules and
integrate them into a module design \cite{sy-harm-flash}.

One can consider different scenarios for modification of the undulator system of the European XFEL in the context of the CW upgrade
of the linac: keeping old undulators, installing new undulators instead of both SASE1 and SASE2, or only instead of one of them.
Note that there is also a potential option of the facility extension with new undulator beamlines.

\section{Conclusion}

In this paper we have considered scenarios for operation of the European XFEL in hard X-ray regime after a possible CW upgrade.
We can conclude that operation in sub-\AA ngstrom regime is possible even at a conservatively assumed beam energy of 7 GeV
provided that bright enough electron beams are available. Recent achievements in technology of high-brightness CW injectors
support this assumption, although further studies on optimization of compression and transport of these beams are required.
Recent tests of prototypes of XFEL cryomodules indicate a possibility that a higher electron energy (about 9 GeV) can be achieved
in CW mode what will make sub-\AA ngstrom operation of the European XFEL in this mode much easier. Correspondingly, operation
in LP mode with a high duty factor will bring the energy into a very comfortable range, above 10 GeV.

We have not discussed the properties of the FEL beam in the case of a reduced energy of the linac. As we have already
mentioned, such a reduction would result in a smaller peak power (and peak brilliance). Energy dependence of output
parameters of the optimized SASE FEL can be estimated with the help of the results of Ref.~\cite{coh-prop}. After choice
of operation points and a detailed optimization of electron beam parameters it will be possible to perform
detailed calculations of the main properties of the FEL radiation in the same way as it was done in \cite{xfel-prop}.


\begin{thebibliography}{99}


\bibitem{flash}
W.~Ackermann et al., Nature Photonics {\bf 1}(2007)336

\bibitem{lcls}
P.~Emma et al., Nature Photonics {\bf 4}(2010)641

\bibitem{sacla}
T.~Ishikawa et al.,
Nature Photonics {\bf 6}(2012)540–544


\bibitem{ks-sase}
A.M.~Kondratenko and E.L.~Saldin, Part. Accelerators {\bf 10}(1980)207

\bibitem{euro-xfel-tdr}
M. Altarelli et al. (Eds.),
XFEL: The European X-Ray Free-Electron Laser. Technical Design Report,
Preprint DESY 2006-097, DESY, Hamburg, 2006 (see also http://xfel.desy.de).

\bibitem{detector}
A. Koch et al., J. Phys.: Conf. Ser. {\bf 425}(2013)062013

\bibitem{brinkmann}
R.~Brinkmann,
Proc. of the LINAC2004 Conference, L\"ubeck, Germany, MO102, [http://www.jacow.org]

\bibitem{jacek}
J. Sekutowicz et al.,
Proc. of IPAC2013 Conference, Shanghai, China, p. 1164,
[http://www.jacow.org]


\bibitem{jacek-fel14}
J. Sekutowicz et al.,
Proc. of the FEL2013 Conference, New York, USA, p. 189, [http://www.jacow.org]

\bibitem{jacek-det}
J. Sekutowicz et al., "Duty factor upgrade of the XFEL linac", to be published.


\bibitem{tesla}
R.~Brinkmann et al. (Eds.),
TESLA technical design report, Part II: Accelerator.
Preprint DESY 2001-11, Hamburg, 2001.


\bibitem{cornell}
C.~Gulliford et al.,
Phys. Rev. ST-AB {\bf 16}(2013)073401


\bibitem{srfgun-fel14}
S.A.~Belomestnykh,
Proc. of the FEL2013 Conference, New York, USA, p. 176, [http://www.jacow.org]


\bibitem{srfgun-ipac14}
M.~Schmeisser et al.,
Proc. of IPAC2013 Conference, San Sebastian, Spain, p. 3146,
[http://www.jacow.org]


\bibitem{berkley-gun}
C.F.~Papadopoulos et al.,
Proc. of the FEL2013 Conference, New York, USA, p. 391, [http://www.jacow.org]



\bibitem{vogel}
V.~Vogel et al.,
Proc. of IPAC2011 Conference, Shanghai, China, p. 282,
[http://www.jacow.org]


\bibitem{jacek-privat}
J. Sekutowicz, private communication.


\bibitem{murphy}
J.B.~Murphy, C.~Pellegrini and R.~Bonifacio,
Opt. Commun. {\bf 53}(1985)197

\bibitem{hg-2}
R.~Bonifacio, L.~De~Salvo, and P.~Pierini,
Nucl. Instr. Meth. A293(1990)627.

\bibitem{kim-1}
Z. Huang and K. Kim, Phys. Rev. E, 62(2000)7295.

\bibitem{mcneil}
B.W.J.~McNeil et al., Phy. Rev. Lett. 96, 084801 (2006)

\bibitem{parisi}
G.~Parisi et al.,
Proc. of the FEL2005 Conference, Stanford, California, USA, p. 187, [http://www.jacow.org]

\bibitem{sy-harm}
E.A.~Schneidmiller and M.V.~Yurkov,
Phys. Rev. ST-AB {\bf 15}(2012)080702

\bibitem{sy-harm-flash}
E.A.~Schneidmiller and M.V.~Yurkov,
Nucl. Instr. Meth. A717(2013)37.

\bibitem{huang}
Z.~Huang et al., private communication

\bibitem{suppression}
E.A.~Schneidmiller and M.V.~Yurkov,
"Suppression of the fundamental wavelength for a successful harmonic lasing in X-ray FELs", to be published

\bibitem{hlss}
E.A.~Schneidmiller and M.V.~Yurkov,
Proc. of the FEL2013 Conference, New York, USA, p. 700, [http://www.jacow.org]

\bibitem{psase}
D.~Xiang et al.,
Phys. Rev. ST-AB {\bf 16}(2013)010703

\bibitem{pizdase}
S.~Serkez et al., preprint DESY-13-135, Aug. 2013

\bibitem{coh-prop}
E.L.~Saldin, E.A.~Schneidmiller and M.V.~Yurkov,
New Journal of Physics {\bf 12}(2010)035010

\bibitem{xfel-prop}
E.A.~Schneidmiller and M.V.~Yurkov,
report DESY 11-152, Hamburg, September 2011


\end{thebibliography}
\end{document}